\documentclass[twocolumn]{aastex7}

\received{\today}
\revised{\today}
\accepted{\today}

\usepackage{amsmath}	
\usepackage{amssymb}
\usepackage[abs]{overpic}
\usepackage{xcolor,varwidth}
\usepackage[normalem]{ulem}

\newcommand{\rg}{M}

\newcommand{\OmegaH}{\Omega_{\rm H}}
\newcommand{\phibh}{\phi_{\rm H}}

\shorttitle{Energy Extraction from Black Holes}
\shortauthors{Chatterjee et al.}

\begin{document}

\title{On the Universality of Energy Extraction from Black Hole Spacetimes}

\author[orcid=0000-0002-2825-3590]{Koushik Chatterjee}
\affiliation{Department of Physics, University of Maryland, 7901 Regents Drive, College Park, MD 20742, USA}
\affiliation{Black Hole Initiative at Harvard University, 20 Garden Street, Cambridge, MA 02138, USA}
\affiliation{Harvard-Smithsonian Center for Astrophysics, 60 Garden Street, Cambridge, MA 02138, USA}
\email[show]{kchatt@umd.edu}  

\author{Ziri Younsi}
\affiliation{Mullard Space Science Laboratory, University College London, Holmbury St.~Mary, Dorking, Surrey, RH5 6NT, UK}
\email{z.younsi@ucl.ac.uk}

\author{Prashant Kocherlakota}
\affiliation{Black Hole Initiative at Harvard University, 20 Garden Street, Cambridge, MA 02138, USA}
\affiliation{Harvard-Smithsonian Center for Astrophysics, 60 Garden Street, Cambridge, MA 02138, USA}
\email{prashant.kocherlakota@cfa.harvard.edu}

\author{Ramesh Narayan}
\affiliation{Black Hole Initiative at Harvard University, 20 Garden Street, Cambridge, MA 02138, USA}
\affiliation{Harvard-Smithsonian Center for Astrophysics, 60 Garden Street, Cambridge, MA 02138, USA}
\email{rnarayan@cfa.harvard.edu}
 
\begin{abstract}

The launching of astrophysical jets provides the most compelling observational evidence for direct extraction of black hole (BH) spin energy via the Blandford-Znajek (BZ) mechanism. Whilst it is known that spinning Kerr BHs within general relativity (GR) follow the BZ jet power relation, the nature of BH energy extraction in general theories of gravity has not been adequately addressed.
This study performs the first comprehensive investigation of the BZ jet power relation by utilizing a generalized BH spacetime geometry which describes parametric deviations from the Kerr metric of GR, yet recovers the Kerr metric in the limit that all deviation parameters vanish.
Through performing and analyzing an extensive suite of three-dimensional covariant magnetohydrodynamics (MHD) simulations of magnetized gas accretion onto these generalized BH spacetimes we find that the BZ jet power relation still holds, in some instances yielding jet powers far in excess of what can be produced by even extremal Kerr BHs.
It is shown that independent variation of the frame-dragging rate of the BH can enhance or suppress the effects of BH spin, and by extension of frame-dragging.
This variation greatly enhances or suppresses the observed jet power and underlying photon ring image asymmetry, introducing a previously unexplored yet important degeneracy in BH parameter inference. Finally we show that sufficiently accurate measurements of the jet power, accretion rate and photon ring properties from supermassive BHs can potentially break this degeneracy, highlighting the need of independent investigations of BH frame-dragging from observations. 

\end{abstract}


\section{Introduction} \label{sec:intro}

Astrophysical jets are an ubiquitous phenomenon in our Universe.
From stellar-mass to supermassive black hole (BH) scales, relativistic outflows of plasma are observed to emanate from regions a few gravitational radii away from the central compact object and extend in excess of several millions to even several billions of gravitational radii, traversing galactic scales.
As jets propagate outwards, they interact with and heat up the surrounding interstellar medium gas by depositing their magnetic and kinetic energy, seeding large-scale turbulent eddies \citep{McNamara:2012, bourne_arepo2017, Cielo:2017}.
Due to their ability to traverse and remain collimated across such large distances at relativistic velocities, astrophysical jets effectively act as energy conduits, transferring their kinetic (and potentially magnetic) energy to the interstellar medium.
They are therefore major contributors to the feedback cycle in active galactic nuclei (AGN) and their energy extraction and deposition mechanisms regulate star formation and galactic evolution \citep{SilkRees98, Magorrian98, Fabian2012, har18feed}. 

The mechanism considered most likely to govern the launching of powerful jets is the Blandford-Znajek (BZ) mechanism \citep{bz77}.
Broadly speaking, within the BZ mechanism framework, the BH forces the magnetic field to co-rotate with the spin direction via the frame-dragging effect, twisting the field lines in the azimuthal direction.
This twisting of magnetic field lines produces an outward magnetic pressure that eventually launches the jet.
Since the twisting of field lines by frame-dragging can be thought of ``work'' done by the BH, there is an effective extraction of the BH's rotational energy.
The BZ outflow efficiency $\eta_{\rm BZ}$, i.e., the ratio of the jet power output and the inflowing power due to gas accretion, can be expressed in terms of the normalized magnetic flux ($\phibh$) measured at the event horizon radius ($r_{\rm H}$), and the horizon angular frequency $\OmegaH:=ac/(2r_{\rm H})$ for a Kerr BH of spin ``$a$'' \citep{bz77,mck04,tch10a}: $\eta_{\rm BZ}\propto \phibh^2 \OmegaH^2$, where $c$ is the speed of light. Hereafter, we assume natural units $G=c=1$ for our calculations, with the characteristic length- and time-scales given by the BH mass $M$. 

General-relativistic magnetohydrodynamics (GRMHD) simulations have been instrumental in shaping our understanding of BH accretion physics, steadily growing in popularity due to their ability to describe the spatio-temporal evolution of the complex gas and magnetic field dynamics for a wide variety of accreting systems.
In this work, we begin with a hydrodynamic torus of weakly magnetized gas, and then introduce a perturbation in the gas properties to trigger disk instabilities and finally, accretion onto the central BH.

\begin{figure*}
    \centering
    \includegraphics[width=\textwidth,trim= 0pt 0pt 0pt 0pt, clip]{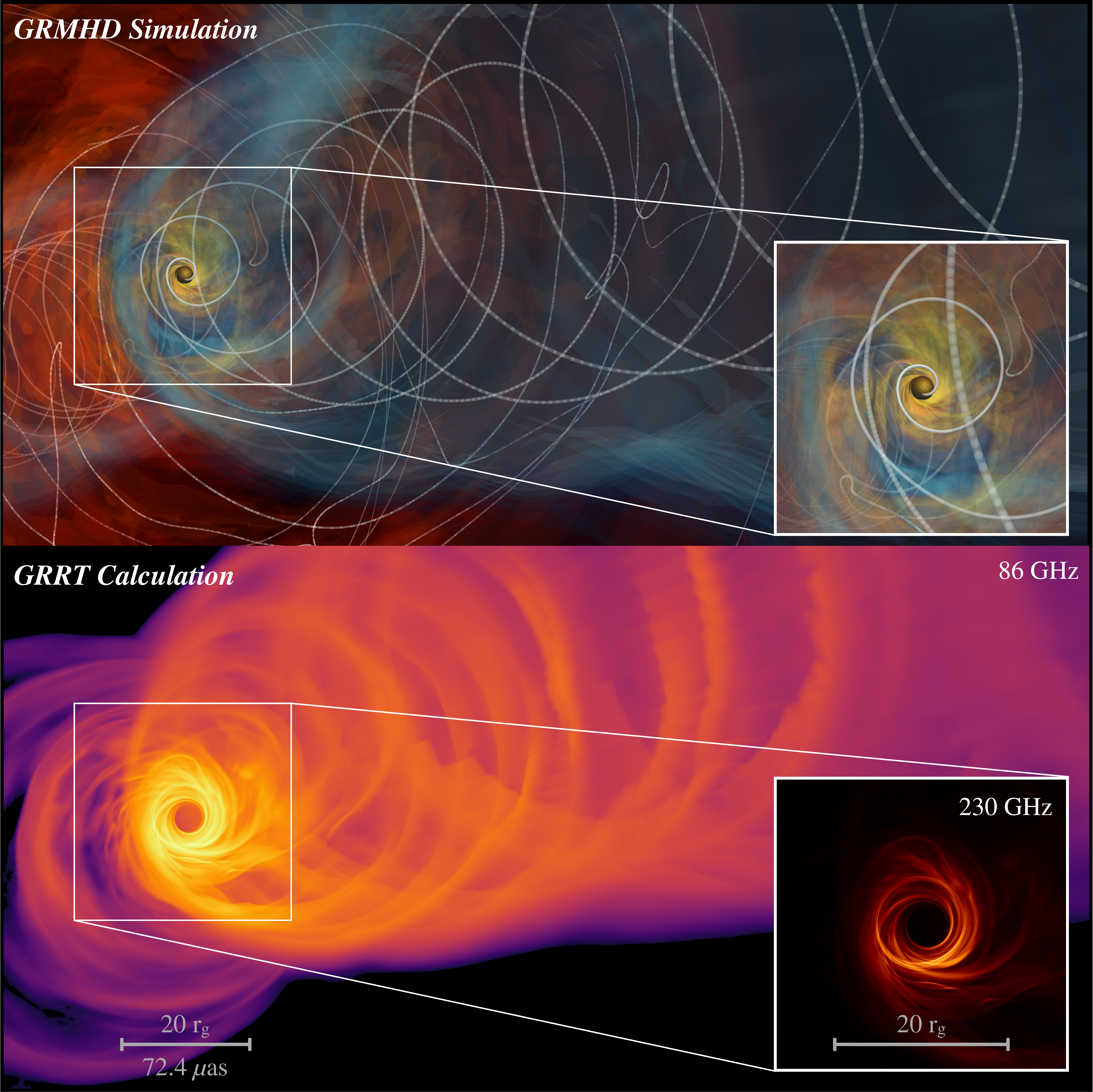}
    \caption{
    Top panel: Snapshot 3D rendering of a BH accreting gas and launching magnetized jets (GRMHD simulations of a JP BH with $\alpha_{22}=2$ and $a=0.9$, see text), together with a zoom-in view of the vicinity of the event horizon (inset).
    The diffuse material comprising the large-scale relativistic jet is shown in blue, with the extended denser accretion disk material shown in red and yellow.
    Tightly-wound magnetic field within the jet and counter-jet is indicated by the gray lines, with the BH event horizon denoted by the central dark sphere.
    Bottom panel: Corresponding ray-traced GRRT image of the upper panel's GRMHD snapshot as viewed at $86$~GHz (logarithmic color scale), together with the inset zoom-in view of the event horizon-scale structure, as viewed at the Event Horizon Telescope (EHT) frequency of $230$~GHz (linear color scale) for M87$^*$.
    }
    \label{fig:visual}
\end{figure*}

As time evolves, three distinct regions develop: a dense equatorial accretion flow with turbulent magnetic fields, a gas-rich, slow-moving wind flowing out from the disk and a near-vacuum jet, filled with helical magnetic fields, moving out with relativistic speeds in the direction of the BH spin axis. Of these three structures, the jet is closest to the BZ solution that assumes force-free electrodynamics. Thus, we expect that BHs where the jet dominates the total power output should, in principle, match the BZ expectation. Figure~\ref{fig:visual} (upper panel) presents a visualization of a BH accretion disk and its large-scale jet. The bottom panel shows our expected observed image of the same jet, exhibiting distinct helical features and hints of limb-brightening in the 86 GHz jet, and the horizon-scale photon ring at 230 GHz, a prediction of BH frame-dragging.

In the limit of a saturated BH magnetosphere, the accretion flow transitions to a magnetically arrested (MAD) state. MADs with highly spinning BHs have distinctive properties such as powerful jets with outflow efficiencies $\eta\gtrsim 100\%$, large enough to cause BH spindown \citep{tch12proc,Narayan:2022,Chatterjee:2022}. MADs also exhibit quasi-periodic magnetic explosions that are thought to be the origin of high-energy flares in the X-ray band reaching up to TeV gamma-rays \citep{Ripperda2022, Hakobyan2023, Zhdankin2023}. The presence of strong vertical magnetic fields near the BH event horizon in MADs also makes them a favorable candidate for explaining the resolved polarization structure of M87* \citep{EHT_M87_2019_PaperVIII}. As MADs describe the theoretical upper limit for the horizon magnetic flux, they are arguably the best candidates for investigating energy extraction from BHs via the BZ mechanism.
Thus we opt for standard MAD initial conditions in this study \citep[][]{EHT_M87_2019_PaperV,EHT_SgrA_2022_PaperV,Chatterjee:2022}.

\begin{table}[htbp!]
\centering
\vspace{10pt}
\renewcommand{\arraystretch}{1.3}
\begin{tabular}{| c | c | c |}
\hline
{\bf Metric} & {\bf Non-Kerr value} & {\bf BH spin parameter}\\
\hline
Kerr & -- & $0$, $\pm 0.3$, $\pm 0.5$, $\pm 0.7$, \\ 
& & $\pm 0.9$, $\pm 0.94$, $\pm 0.998$ \\
\hline
JP $\alpha_{13}$ & $-2$ & $0$, $\pm 0.3$, $\pm 0.5$\\
JP $\alpha_{13}$ & $+2$ & $0$, $\pm 0.3$, $\pm 0.5$, $\pm 0.7$\\
JP $\alpha_{13}$ & $+6$ & $0$, $\pm 0.3$, $\pm 0.5$, $\pm 0.7$\\
\hline
JP $\alpha_{22}$ & $-2$ & $0$, $\pm 0.3$, $\pm 0.5$, $\pm 0.7$, $0.9$\\
JP $\alpha_{22}$ & $+2$ & $0$, $\pm 0.3$, $\pm 0.5$, $\pm 0.7$, $0.9$\\
JP $\alpha_{22}$ & $+5$ & $0$, $\pm 0.3$, $\pm 0.5$, $\pm 0.7$\\
JP $\alpha_{22}$ & $+8$ & $\pm 0.3$, $\pm 0.5$\\
\hline
JP $\epsilon_{3}$ & $-3$ & $0$, $\pm 0.3$, $\pm 0.5$\\
JP $\epsilon_{3}$ & $+6$ & $0$, $\pm 0.3$, $\pm 0.5$\\
\hline
\end{tabular}
\caption{List of GRMHD simulations performed in this work, together with their respective spin and non-Kerr parameter values.} 
\label{tab:models}
\end{table}

\section{Methods} \label{sec:methods}

We use the GPU-accelerated GRMHD code \texttt{H-AMR} \citep{liska_hamr:2022} which solves the GRMHD equations in a fixed but arbitrary background spacetime. We assume logarithmic spherical polar in-going Kerr-Schild coordinates ($t$, $\log r$, $\theta$, $\varphi$) and natural units, i.e. $G=c=1$, which normalizes the gravitational radius $r_{\rm g}=GM_{\rm BH}/c^2$ to the BH mass $M$. 
Our simulations have an effective grid resolution of $N_r\times N_{\theta}\times N_{\varphi}=348\times 240\times256$.
The grid extends from $r/\rg\in (1.1525,10000)$, $\theta\in (0,\pi)$ and $\varphi\in (0,2\pi)$. In order to speed up our simulations, we de-refine the $\varphi-$resolution by a factor of 4 near the polar axis.  We adopt outflowing radial boundary conditions (BCs), transmissive polar BCs and periodic BCs in the $\varphi-$direction \citep{Liska:18a,chatterjee2019,liska_hamr:2022}. We initialize the disk in the form of an equilibrium hydrodynamic torus \citep{Fishbone:76} around a BH with a fiducial spin value of $a=0.9375$. We set up a standard MAD magnetic field configuration \citep[e.g.,][]{Chatterjee:2022}. For our gas thermodynamics, we assume an ideal gas equation of state with an adiabatic index of $13/9$. The initial magnetic field strength is normalized by setting the initial maximum gas-to-magnetic pressure ratio to 100. For tackling the evacuated region in the jet funnel, we adopt the density floor injection scheme of \citet{Ressler:17} when the magnetization exceeds 20. 

The expected jet power efficiency from the BZ mechanism is given as, 
\begin{equation}
    \eta_{\rm BZ} = \frac{P_{\rm BZ}}{\dot{Mc^2}} = \frac{k}{4\pi} \, \phibh^2 \, \OmegaH^2 \,,
    \label{eqn:BZ}
\end{equation}
\noindent where the constant $k$ depends on the parabolicity of the poloidal magnetic field ($k=0.054$ for near-monopolar or purely radial magnetic fields and $0.044$ for parabolic fields). Hereafter, we assume natural units $G=c=1$ for our calculations, with the characteristic length- and time-scales given by the BH mass $M$. 

The dimensionless magnetic flux is defined as:
\begin{equation}
    \phibh := \frac{\sqrt{4\pi}}{2\sqrt{\dot{M}}}\iint |B^{r}| \, \sqrt{-g}\, d\theta \, d\varphi \,,
    \label{eqn:phibh}
\end{equation}
where the integral is calculated at the event horizon radius $r=r_{\rm H}:=\rg (1+\sqrt{1-a^2})$ (since the chosen JP parameters do not change the horizon size).
In eqn.~\eqref{eqn:phibh}, $B^r$ is the radial magnetic field component, $g$ is the metric determinant, and $\dot{M}$ is the shell-integrated mass accretion rate, $\dot{M}(r) = -\iint \rho u^r \, \! \sqrt{-g}\, d\theta \, d\varphi$, calculated at $r=5\,\rg$ in order to avoid contamination from density floors.

The outflow efficiency from the GRMHD simulations is given by:
\begin{equation}
    \eta=\frac{P}{\dot{M}c^2}=\frac{\dot{M}c^2-\dot{E}}{\dot{M}c^2} \,.
    \label{eqn:eta}
\end{equation}

Here $P$ is the outflow power and $\dot{E}=\iint T^r_t \, \! \sqrt{-g}\, d\theta \, d\varphi$ is the total energy flux in the radial direction. The $(r, t)$ component of the stress-energy tensor is expressed as $T^r_t=(\rho + \gamma_{\rm ad} U_{\rm g}+b^2)u^ru_t - b^rb_t$. Here, $\rho$ represents the gas density, $\gamma_{\rm ad}$ its adiabatic index, $U_{\rm g}$ its internal energy, $u^\mu$ its four-velocity, and $b^\mu$ the magnetic four-vector. 
We calculate the outflow efficiency at $r=5\,\rg$. 
All of these simulations are evolved to $t=25000M$ to make sure that the inner $30~M$ are in inflow-outflow equilibrium. Here we only discuss quantities time-averaged between $20000~M-25000~M$.

For imaging, we use the GRRT raytracing code \texttt{BHOSS} \citep{younsi_2012,younsi_2016,younsi_2019_polarizedbhoss,Younsi2023}.
We assume that the emission mechanism is thermal synchrotron radiation for a (relativistic) Maxwell-J\"uttner electron population.
The electron temperature is calculated from the temperature of the ions via the ion-to-electron temperature ratio $T_{\rm i}/T_{\rm e} := (R_{\rm low} + \beta^2\, R_{\rm high})/(1 + \beta^2)$ \citep{Moscibrodzka_2016}, where the local plasma-$\beta$ parameter, together with the dimensionless parameters $R_{\rm low}$ and $R_{\rm high}$, control the coupling between ions and electrons.
We choose $R_{\rm low}=1$ and $R_{\rm high}=40$ in all models, consistent with parameter fits of recent polarimetric observations of M87* \citep{EHT_M87_2019_PaperVIII}.
The images presented represent the average over the time interval $20000$~M--$25000$~M, where we have assumed the mass and distance of M87* to be $6.2\times 10^{9}\,M_{\odot}$ and $16.9\,{\rm Mpc}$, respectively.
We have scaled the black hole mass accretion rates from code units to physical units such that we obtain M87*'s time-averaged $230~{\rm GHz}$ flux of $0.5\,{\rm Jy}$ \citep{EHT_M87_2019_PaperV}.

In order to explore deviations from the Kerr metric and GR, we consider the Johannsen-Psaltis (JP) metric \citep[][]{Johannsen:2011,Johannsen:2013_JP,Johannsen:2013_JP_rings}.
The JP metric describes the exterior solution of a rapidly spinning BH and introduces parametric deviations to the Kerr metric via adjustable dimensionless deviation parameters (see Appendix~\ref{sec:metric}). At lowest order there are four such parameters: $\epsilon_{3},\,\alpha_{13},\,\alpha_{22},\,\alpha_{52}$, of which we always set $\alpha_{52}=0$ since this parameter affects only the $g_{rr}$ metric component and does not alter the angular frequency or the innermost stable circular orbit (ISCO) of particles, nor the unstable photon orbit radius and therefore the observed photon ring image.
The geodesic integrability and topological structure of the JP metric are well understood, enabling the study of rapidly spinning BHs that cannot be
described by the Kerr solution, nor be admitted as a solution of the vacuum Einstein field equations of GR.

\begin{figure*}
    \centering
    \includegraphics[width=\textwidth,trim= 110pt 0pt 140pt 0pt, clip]{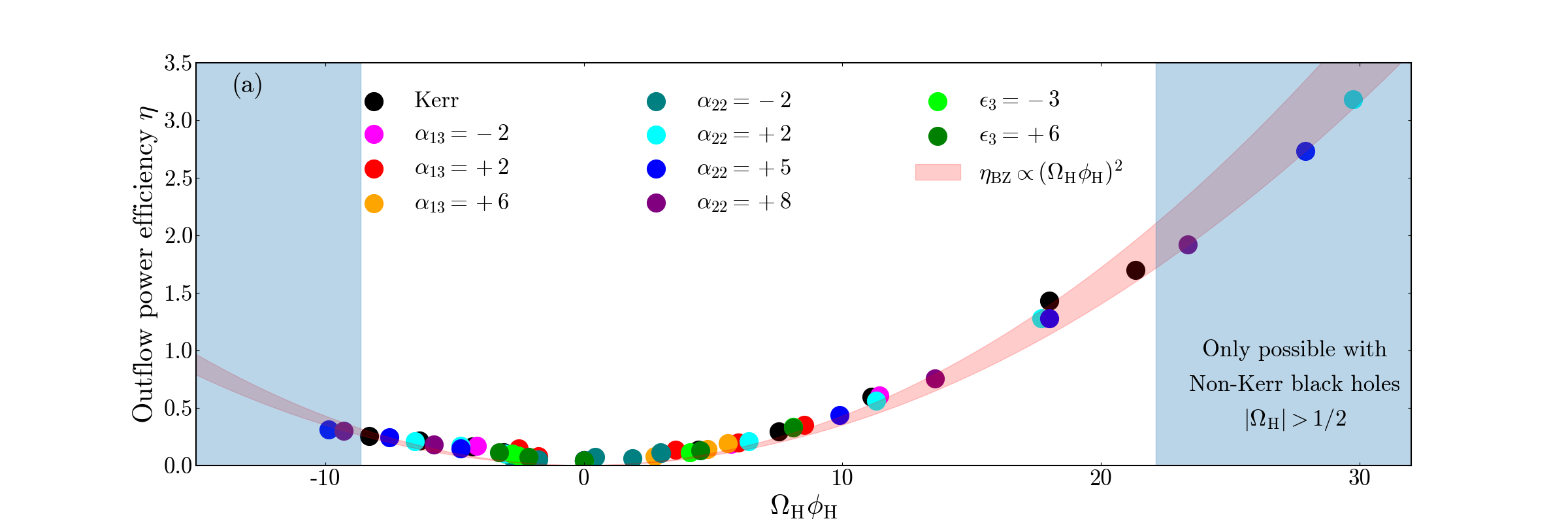}
    \includegraphics[width=\textwidth,trim= 110pt 0pt 140pt 0pt, clip]{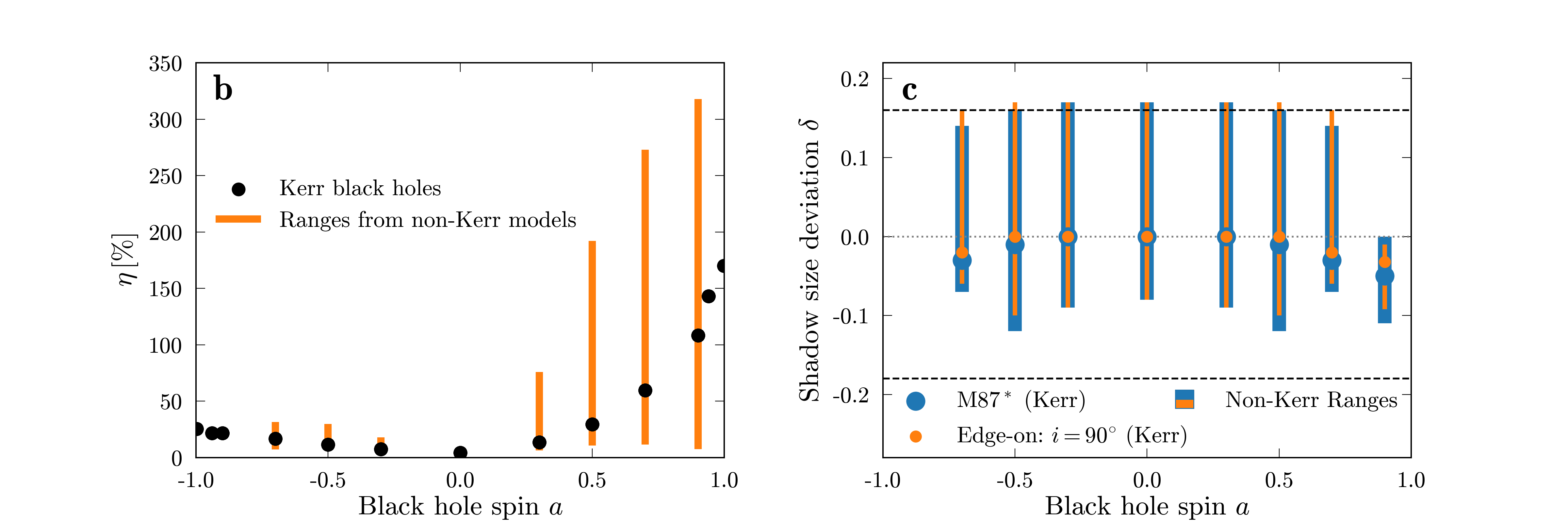}
    \caption{Panel {\bf a} demonstrates that the BZ mechanism accurately describes the outflow power for arbitrary non-Kerr BHs out to large horizon angular frequencies beyond even extremal Kerr BHs.
    This implies that the BZ power is a fundamental property of spinning BHs. The red shaded region indicates the predicted BZ power over a range of magnetic field shapes: $0.044 \leq k\leq 0.054$ from eqn.~\ref{eqn:BZ}.
    Panel {\bf b} shows that the outflow power can assume a large range of values given a particular BH spin within our model set.
    Panel {\bf c} is similar to panel {\bf b}, but now presents the deviation in mean BH shadow diameter with respect to the Schwarzschild value of $d_{\rm sh, Schw}=6\sqrt{3}~M$, hereafter $\delta :=d_{\rm sh}/d_{\rm sh, Schw}-1$.
    We also show the 2017 EHT measurements of M87* of $\delta=-0.01\pm0.17$ \citep{EHTC+2019f, Psaltis+2020, Kocherlakota+2021}, delineated as dashed lines.
    These plots illustrate the inherent degeneracies in BH spin inference from jet power estimates and shadow image size measurements when one relaxes the assumption that the spacetime geometry is described by a Kerr BH. See Sec.~\ref{sec:conclusion} for more discussion on these degeneracies.
    }
    \label{fig:BZ}
\end{figure*}

In this work we independently adjust the remaining three parameters: $\epsilon_{3}$ only weakly alters the Kerr ISCO radius, $\alpha_{13}$ is $\mathcal{O}\left( r^{-3} \right)$ and mildly alters the ISCO radius, and $\alpha_{22}$ is $\mathcal{O}\left( r^{-2} \right)$ and significantly alters the ISCO radius.
Regarding the photon ring radius, $\epsilon_{3}$ has no effect, $\alpha_{13}$ alters its Kerr value and ``circularizes'' the observed ring, and $\alpha_{22}$ also alters the ring radius but serves to enhance asymmetry.
This latter point may be understood as $\alpha_{22}$ adjusting the frame-dragging rate of the BH \citep{Younsi2023}. The horizon angular frequency for the JP metric depends not only on the spin, but also on $\alpha_{13}$ and $\alpha_{22}$ in the following manner:
\begin{equation}
    \OmegaH^{\rm JP} = \frac{a}{2r_{\rm H}} \frac{A_2}{A_1}{\Bigg |}_{\rm r=r_{\rm H}} \,,
    \label{eqn:omegaH}
\end{equation}
where $A_1 \equiv 1+\alpha_{13}/r^3$ and $A_2 \equiv 1+\alpha_{22}/r^2$. Thus, increasing $\alpha_{22}$ drives up the horizon angular frequency by enhancing frame-dragging while increasing $\alpha_{13}$ serves to suppress frame dragging, in effect decreasing $\OmegaH$. Here, $r_{\rm H}$ denotes the coordinate radius of the event horizon and is determined purely by the BH mass and spin, $r_{\rm H} = M + \sqrt{M^2 - a^2}$. One therefore anticipates that large horizon angular frequencies (i.e., large values of $\alpha_{22}$) will enhance energy extraction from the BH and yield much more powerful jets than even an extremal Kerr BH.

\section{Results} \label{sec:results}

We perform a suite of 69 distinct 3D covariant MHD simulations including general relativity and parametric deviations from it with the JP metric. For brevity, we will refer to all simulations as simply ``GRMHD''.
Of these simulation, 13 employ the Kerr metric with $a=(0,\,\pm 0.3,\, \pm 0.5,\, \pm 0.7,\, \pm 0.9, \pm 0.94,\, \pm 0.998)$, 19 simulations account for deviations solely in $\alpha_{13}$, 27 simulations for deviations solely in $\alpha_{22}$, and finally 10 simulations for deviations solely in $\epsilon_3$ (see Table~\ref{tab:models}). 

We focus on the calculation of the dimensionless magnetic flux on the BH event horizon $\phibh$ and the outflow efficiency $\eta$, comprising the contributions of both the jet and disk wind to outward energy transfer. We also capture the outflow of energy via magnetic flux eruptions that are mediated via magnetic reconnection \citep{Ripperda2022}, the properties of which depend on numerical grid resolution. However, studies of numerical resolution convergence of MADs by \citet{Salas:2024} found that the time-averaged value of $\phibh$ and $\eta$ are broadly consistent between MAD simulations of varying grid resolution. For measuring $\eta$, we calculate the total radial energy flux and the mass accretion rate directly from our simulations, and thus, do not assume any apriori jet launching process. We refer to time-averaged quantities in this Letter (see Appendix \ref{sec:fluxes} for a discussion on time evolution of our simulations).

Figure~\ref{fig:BZ}a shows a comparison between the time-averaged total outflowing power efficiency $\eta$ from our simulations with the corresponding expected BZ power. We find that the outflow power follows the BZ formula for both Kerr and non-Kerr models, even at large values of $\OmegaH\phibh$, i.e., when the jet dominates the total outflow. It is noteworthy that the maximal $|\OmegaH|=1/2$ for Kerr BHs, whereas the $\alpha_{22}\neq0$ models can reach larger horizon frequencies for the same BH spin. The accretion flow for our MAD models exhibit a large range of horizon magnetic fluxes and outflow powers, all of which follow the predicted BZ power-relation (also see Appendix~\ref{sec:fluxes}), which clearly shows that the BZ process is not only the primary energy extraction mechanism for accreting BHs, and is universally applicable for arbitrary spacetimes. 
\begin{figure*}
    \centering
    \includegraphics[width=1\textwidth]{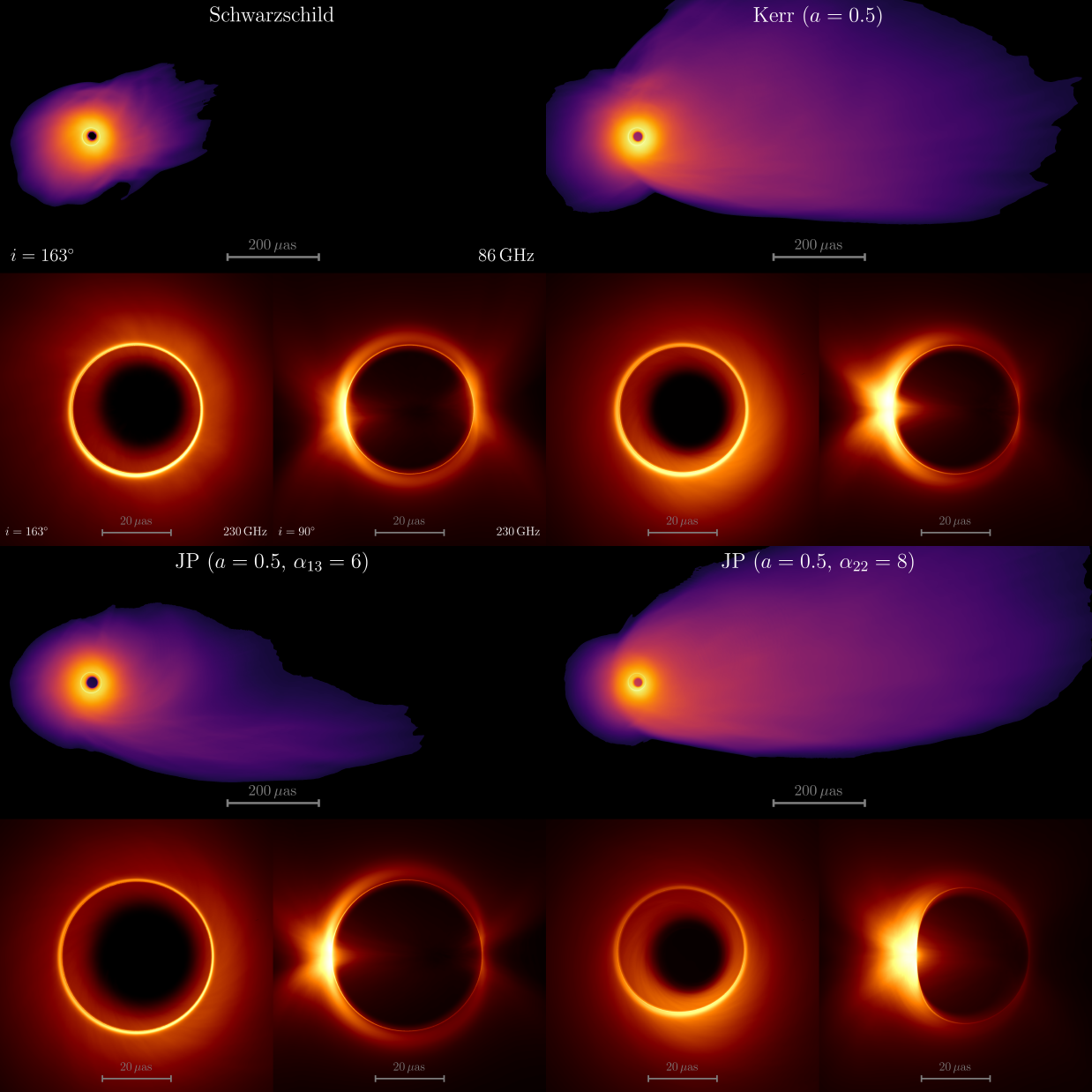}
    \caption{Collection of three-image panel groups showing $86$~GHz M87* images (upper panels) and corresponding 230~GHz BH shadow images (both viewed at $i=163^{\circ}$, bottom left) and as viewed at $i=90^{\circ}$ (bottom right).
    From left to right, top to bottom, trios of panels represent: Schwarzschild BH, Kerr BH with $a=0.5$, JP BH with $\alpha_{13}=6$ and $a=0.5$, and a JP BH with $\alpha_{22}=8$ and $a=0.5$, respectively.
    Differences in jet size and power are evident between Schwarzschild and Kerr BHs at $86$~GHz.
    The JP BH with $\alpha_{13}=6$, in spite of having the same spin as the Kerr BH, has a significantly weaker jet and larger photon ring size.
    By contrast, the JP BH with $\alpha_{22}=8$ possesses a large $\OmegaH$, enhancing the effects of frame dragging, giving rise to a more powerful and extended jet than the Kerr BH with identical spin.
    All panels show time-averaged images over the interval $20,000$~M -- $25,000$~M. The $86$~GHz image color scale is logarithmic and spans 5 orders of magnitude in specific intensity, whereas the $230$~GHz images are plotted on a linear scale and show the full range of specific intensity. All images are normalized to a maximum pixel intensity of unity.
}
    \label{fig:image}
\end{figure*}

Evidently, a degeneracy in $\eta$ arises between BH spin and non-Kerr parameters.
Panel (b) of Figure~\ref{fig:BZ} shows that for a given spin, the range of possible outflow efficiencies for a non-Kerr BH is large, especially for larger values of $a$.
Measuring the jet power may therefore not provide an accurate estimate of BH spin when $\alpha_{22}$ is of large and opposite sign, mimicking the effects of spin.
Caution must therefore be exercised.
For example, consider that for the highly spinning $a=0.9$ BH, a value of $\alpha_{22}=-2$ results in a near-zero horizon frequency and barely produces an outflow, while $\alpha_{22}=2$ exhibits $\eta\approx 330\%$, a factor of $\approx 2$ larger than the Kerr $a=0.998$ model.

To see the effect of introducing non-Kerr parameters on BH images, we performed general-relativistic radiative transfer (GRRT) calculations of a selection of GRMHD simulations with the raytracing code \texttt{BHOSS} \citep{younsi_2012,Younsi2023}, focusing on comparison with an $a=0.5$ Kerr BH model.
Figure~\ref{fig:image} shows the $86$~GHz and $230$~GHz images for M87* parameters: a BH mass of $M_{\rm BH}=6.2\times10^9M_{\odot}$ located at a distance of $D_{\rm BH}=16.9\,$ Mpc (around $55$ million lightyears) viewed at a nearly face-on inclination of $i=163^{\circ}$ \citep{EHTPaperV}.

The $86$~GHz images for BH models with large outflow powers show more extended and distinctly collimated jets.
Usually, non-thermal processes such as magnetic reconnection and shocks in jets \citep{sironi2015} need to be invoked to reproduce large-scale extended jet images that are seen in astronomical observational, e.g., out to sizes $\gtrsim 500~\mu$as for M87* \citep{Fromm:2022}.
We find that when we increase the frame-dragging rate of the BH, we can reach such image sizes for moderate BH spins even without assuming non-thermal radiative processes.
Essentially, we can increase $\alpha_{22}$ to achieve a larger $\OmegaH$, which in turn drives a more efficient and powerful jet (and changes its collimation profile, see Appendix~\ref{sec:collimation}), creating an extended image in a manner akin to how an increase in the BH spin leads to more extended jet images for Kerr BHs \citep[][]{Fromm:2022}.
Since non-Kerr spacetimes permit $\OmegaH>1/2$, jet images can be much more extended than is possible with even an extremal Kerr BH.
On the other hand, we can also decrease $\OmegaH$ with the $\alpha_{13}$ parameter, resulting in a weakly powered jet that appears similar to the Schwarzschild $86$~GHz jet image.
Thus, as seen through $\eta$, the size of the $86$~GHz image is intimately linked to the BH spacetime via the BZ process.
Consequently, observing a large-scale jet does not necessarily imply a rapidly spinning BH, as a small spin combined with a large $\alpha_{22}$ can enhance the spin and hence the amount of frame dragging, yielding a similar image to that produced by a more rapidly spinning Kerr BH. 

One should also exercise caution when estimating spin from BH shadow size measurements.
Panel (c) of Fig.~\ref{fig:BZ} presents the theoretical mean shadow size calculations for Kerr BHs when viewed edge-on (i.e., an observer inclination angle $i=90^{\circ}$) and at $i=163^{\circ}$ (i.e., the inclination angle for M87*).
The filled-in area indicates the range of shadow sizes allowed by our particular set of non-Kerr parameters for a given BH spin.
We find that the variation in the shadow diameters is largest for a face-on observer whereas the asymmetry of the ring is largest for an edge-on observer, due to frame-dragging (see also Fig.~\ref{fig:image}). For the case of M87$^*$, the observer is nearly face-on ($i=163^{\circ}$) and the shadows are effectively circular. The BH models we use here all have similar shadow sizes.

Turning to the $230$~GHz horizon-scale images, the image structure for M87* looks very similar between the models, as expected from panel (c) of Fig.~\ref{fig:BZ}.
We also include the corresponding edge-on inclination cases to demonstrate the maximum deformation of the shadow shape and asymmetry of the image seen in Fig.~\ref{fig:BZ}c.
In particular, as compared to the Kerr edge-on image ($i=90^{\circ}$), the corresponding shadow for $\alpha_{22}=8$ shows a significant enhancement in prolateness and shifts towards the right, mimicking the effects of strong frame-dragging exhibited by rapidly spinning Kerr BHs.
In fact, the horizon angular frequency of a JP BH with spin $a=0.5$ and $\alpha_{22}=8$ is similar to that of a Kerr BH with spin $a=0.992$.
Further, we see that the $\alpha_{13}=6$ image looks remarkably similar to the Schwarzschild image since $\alpha_{13}$ acts to circularize the shadow image's shape, counteracting the effects of spin by reducing $\OmegaH$, accompanied by a small increase in the shadow diameter.
This reaffirms the large degeneracy in horizon-scale image morphologies between Kerr and non-Kerr BHs.  

\begin{figure*}
    \centering
    \includegraphics[width=0.317\textwidth,trim= 0pt 0pt 0pt 0pt, clip]{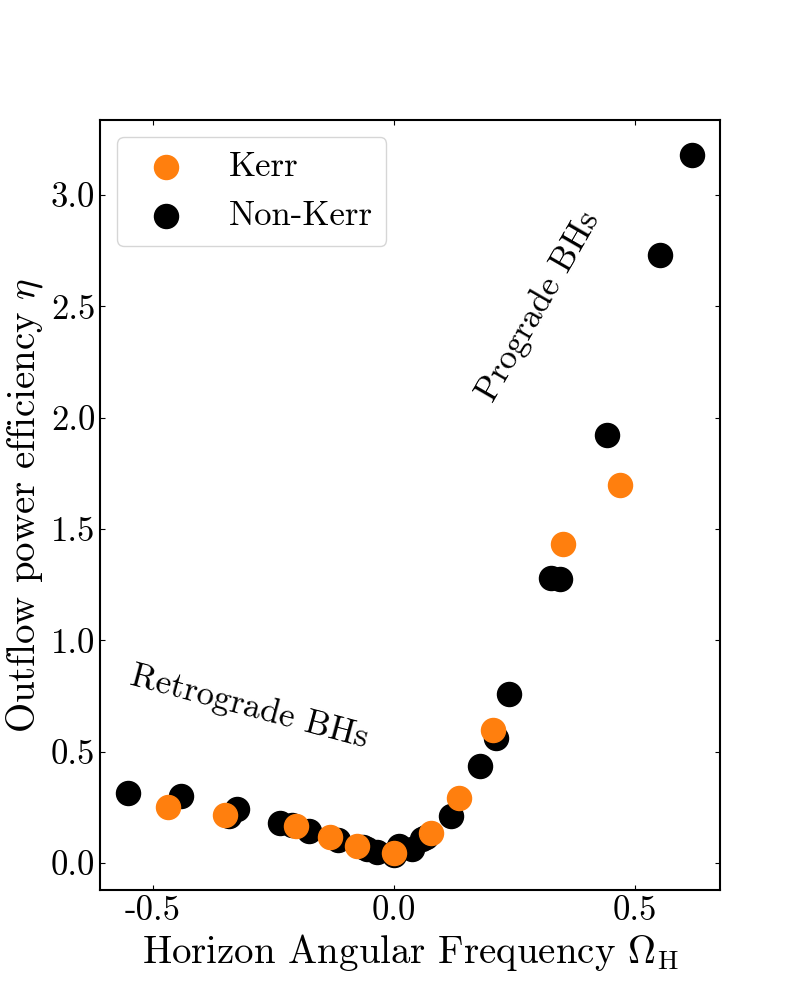}
    \includegraphics[width=0.673\textwidth,trim= 0pt 0pt 0pt 0pt, clip]{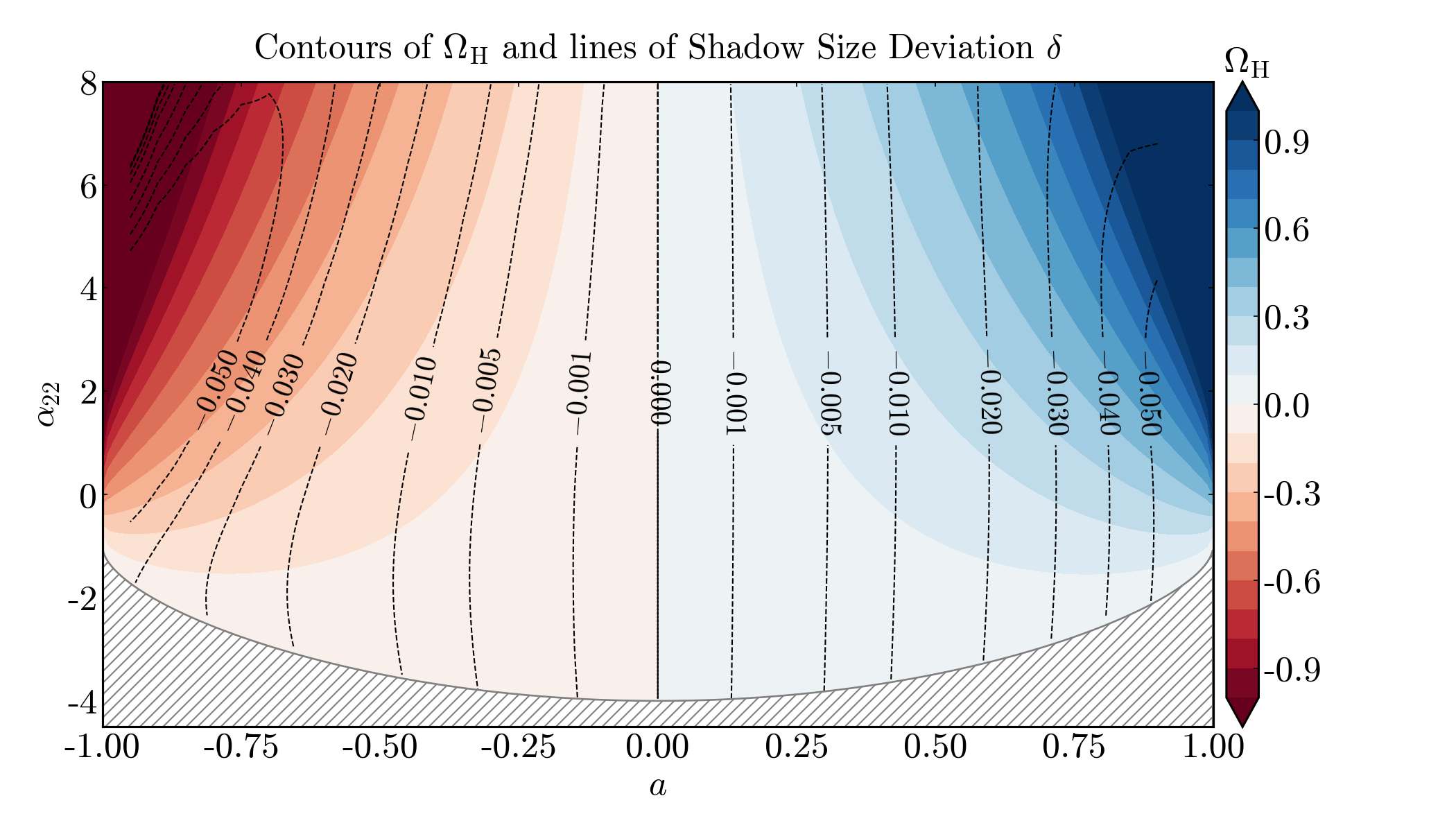}
    \caption{Breaking degeneracies in BH parameters. \textit{Left:} The outflow power $\eta$ depends strongly on the horizon angular frequency $\Omega_{\rm H}$ for the Kerr and Non-Kerr JP $\alpha_{22}$ models. $\eta$ increases monotonically as $|\Omega_{\rm H}|$ increases both for prograde and retrograde BHs. Therefore, it is possible to constrain $|\Omega_{\rm H}|$ using measurements of the outflow efficiency from observations of jetted BHs. \textit{Right:} Contours of $\Omega_{\rm H}$ (in colour) overlaid with contours of the shadow size deviation $\delta$ for JP $\alpha_{22}$ spacetimes. Independent measurements of $\Omega_{\rm H}$ and $\delta$ could place stringent constraints on the black hole spin $a$ and quadrapole moment $\alpha_{22}$. Forbidden solutions for the JP metric are shown by the gray-lined region. 
    }
    \label{fig:degen}
\end{figure*}

\section{Discussion and Conclusion} \label{sec:conclusion}

In this Letter, we present 3D numerical simulations of spinning non-Kerr BHs, building the most extensive library of 3D GRMHD simulations to date. We demonstrate that the Blandford-Znajek mechanism describes relativistic jet launching for a large class of BHs using a parameterized BH metric. We verify that the BZ power efficiency $\eta_{\rm BZ}\propto \phibh^2\OmegaH^2$ describes the outflow power for arbitrary spinning BH spacetimes \citep[also see][]{Chatterjee:2023_KS}. It is interesting to note that the BZ power formula has been extended to higher BH spins in orders of $\OmegaH^2$, out to $\mathcal{O}(\OmegaH^8)$  \citep{tn08,tch10a, Camilloni:2022}, to preserve the non-dependence of the power output on BH spin direction for extremely high Kerr BH spins ($a\gtrsim0.99$), in the form of $\eta_{\rm BZ, ext} = \eta_{\rm BZ} \times f(\OmegaH)$, where $f(\OmegaH)=1+1.38~\OmegaH^2 - 9.2~\OmegaH^4$ \citep{tch10a}. Most semi-analytical work on BZ jet powers focus on such extensions of $\OmegaH$. However, this formulation ignores the effect of BH spin on $\phibh$ as well as the properties of the accretion flow, which becomes an important assumption for self-consistently evolved accreting BHs.
In the magnetically saturated limit (i.e., the MAD limit), the magnetic flux at the event horizon reaches the maximum possible value for a particular disk geometry. Decreasing the geometrical thickness of the accretion flow lowers the corresponding MAD $\phibh$ value \citep{Lowell:2023, Liska:2024}. Hence, for our models, which are applicable for geometrically-thick, sub-Eddington accretion flows such as M87* ($\dot{M}\ll 10^{-3}\dot{M}_{\rm Edd}$) \citep{Chatterjee:2023}, the time-averaged $\phibh$ could be deemed as an upper-limit. We find that $\phibh$ depends strongly on not only the BH spin, verifying results from previous work on Kerr BHs \citep{Narayan:2022}, but also on each non-Kerr deviation parameter. 

Constraining the properties of BHs, in particular their spins, is of significant importance as it provides crucial insights into BH formation and evolution over time. We show that large degeneracies exist in both outflow efficiency as well as shadow and image sizes between Kerr and non-Kerr BHs. Even for the non-Kerr models that exhibit shadow size deviations within the measured bounds from M87*, there is a large change in the jet efficiency, which could affect the 86GHz--230GHz spectral index and polarization signatures especially during flaring events \citep{Gelles:2022, Jia:2023}. Thus, caution should be exercised when inferring the BH spin from current horizon-scale and jet observations. 

A large body of work exists on measuring the BH spin indirectly through modeling disk emission using, e.g., X-ray reflection methods and, more recently, modeling gravitational wave signals \citep[e.g., see review by][]{Reynolds:2019}. 
Further, one can estimate the BH spin by linking the observed radio luminosities of jetted BHs (thereby obtaining a measure of intrinsic jet power) and accretion rate estimates from X-ray observations \citep[such as for M87;][]{russellD2013} or radio polarimetry \citep{Marrone_2007} by assuming the BZ mechanism of jet launching \citep[see references in e.g.,][]{Daly:2019}. 
Some of these methods have been extended to constrain non-Kerr parameters \citep[e.g.,][]{Bambi:2012,Bambi:2017}. 
While most of these works omit detailed specification of the accretion flow properties, particularly for more complex disk geometries, magnetic field strengths, and orientations, they represent major steps towards understanding the BH spin population. 

Resolved near-horizon imaging of supermassive BHs opens a new window to estimate the strength of frame-dragging via direct comparisons with GRMHD-based accretion models \citep{EHT_M87_2019_PaperV, EHT_SgrA_2022_PaperV}. 
Even more promising is the strong dependence of horizon-scale polarization maps on the BH spin for Kerr BH accretion models \citep{Palumbo:2020,EHT_M87_2019_PaperVIII,Emami:2023,Chael_2023,Gelles_2025}, not withstanding the uncertainties due to Faraday effects from the surrounding gas. 
In this Letter we have shown that for the same BH spin, increasing the frame-dragging parameter $\alpha_{22}$ increases the strength of the frame-dragging (i.e., $\Omega_{\rm H}$), which directly affects the horizon-scale polarization structure. Consequently, polarization maps would instead constrain the value of $\Omega_{\rm H}$, resulting in a new independent measurement of frame-dragging.

Figure~\ref{fig:degen} shows that independent, high-precision measurements of $\Omega_{\rm H}$ and the shadow size deviation $\delta$ have the potential to constrain both the BH spin, $a$, as well as any enhancements of frame-dragging via $\alpha_{22}$, particularly for rapidly spinning BHs, thereby providing a path towards constraining the ``Kerr-ness'' of astrophysical BHs. 
While there are currently promising avenues for conducting precision measurements of frame-dragging, of which some examples are provided above, there still remains large uncertainty in the shadow size deviation (e.g., for that of M87$^*$ where the measured $\delta=-0.01\pm0.17$).
Current EHT measurements have difficulty in distinguishing Kerr BH images from BH images in alternative theories of gravity \citep[see][]{Mizuno:2018}.
Repeated observations with the EHT and future space-based VLBI experiments \citep[see, e.g.,][]{Fromm_2021,Roelofs_2021} will sufficiently lower the measurement error in $\delta$ such that estimating $\alpha_{22}$ will become feasible, e.g., through using Fig.~\ref{fig:degen}. Therefore, horizon-scale image structures with future high-angular resolution VLBI, together with accurate independent measurements of the BH mass (such as for M87*) and jet power, have the potential to more stringently constrain the space-time properties of supermassive BHs.

\begin{acknowledgments}
 We thank the anonymous referee for their thoughtful comments and suggestions. KC, PK and RN are supported in part by grants from the Gordon and Betty Moore Foundation and the John Templeton Foundation to the Black Hole Initiative at Harvard University, and by NSF award OISE-1743747. KC additionally acknowledges support by NASA grant 80NSSC22K1054. ZY acknowledges support from a UK Research and Innovation (UKRI) Stephen Hawking Fellowship. This research was enabled by support provided by a INCITE program award PHY129, using resources from the Oak Ridge Leadership Computing Facility, Summit, which is a US Department of Energy office of Science User Facility supported under contract DE-AC05- 00OR22725, as well as Calcul Quebec (http://www.calculquebec.ca) and Compute Canada (http://www.computecanada.ca). This work has made use of NASA's Astrophysics Data System (ADS).
\end{acknowledgments}

\appendix

\section{\label{sec:metric}Parametrized metric for spinning black holes}
As discussed in the Letter, all calculations are performed using a generalized parametrization of black hole (BH) spacetime metrics using the Johannsen-Psaltis metric \citep{Johannsen:2011,Johannsen:2013_JP,Johannsen:2013_JP_rings}.
The line element for the JP metric is expressed in Boyer-Lindquist form as
\begin{equation}
\begin{aligned}
{\rm d}s^{2} = -\frac{\widetilde{\Sigma} \, \mathcal{B}}{\mathcal{F}} \, {\rm d}t^{2} - \frac{2 a \, \widetilde{\Sigma} \, \mathcal{C} \sin^{2} \theta}{\mathcal{F}} \,{\rm d}t \, {\rm d}\phi + \frac{\widetilde{\Sigma}}{\Delta A_{ 5}} \, {\rm d}r^{2} + \widetilde{\Sigma} \, {\rm d}\theta^{2} + \frac{\widetilde{\Sigma} \, \mathcal{D} \sin^{2}\theta}{\mathcal{F}} \, {\rm d}\phi^{2} \,,
\end{aligned}
\end{equation}
where
\begin{subequations}
\begin{eqnarray}
\mathcal{B} &\equiv& \Delta - a^{2}A_{ 2}^{2}\sin^{2}\theta \,, \\
\mathcal{C} &\equiv& \left( r^{2} + a^{2} \right) A_{1} A_{2} - \Delta \,, \\
\mathcal{D} &\equiv& \left( r^{2} + a^{2} \right)^{2} A_{1}^{2} - a^{2} \Delta \sin^{2}\theta \,, \\
\mathcal{F} &\equiv& \left[ \left(r^{2} + a^{2}\right) A_{1} - a^{2} A_{2}\sin^{2} \theta \right]^{2} \,,
\end{eqnarray}
\end{subequations}
and where $\Sigma \equiv r^{2} + a^{2} \cos^{2} \theta$ and $\Delta \equiv r^{2} - 2Mr + a^{2}$, as for Kerr, and terms depending on the Kerr metric deformation parameters are defined as
\begin{subequations}
\begin{eqnarray}
\widetilde{\Sigma} &\equiv& \Sigma + M^{2} \, \sum_{n=3}^{\infty} \epsilon_{n} \, \left(\frac{M}{r}\right)^{n-2} \,, \\
A_{1} &\equiv& 1 + \sum_{n=3}^{\infty} \alpha_{1 n} \left( \frac{M}{r} \right)^{n} \,, \\
A_{2} &\equiv& 1 + \sum_{n=2}^{\infty} \alpha_{2 n} \left( \frac{M}{r} \right)^{n} \,, \\
A_{5} &\equiv& 1 + \sum_{n=2}^{\infty} \alpha_{5 n} \left( \frac{M}{r} \right)^{n} \,.
\end{eqnarray}
\end{subequations}
The above metric deformation parameters are dimensionless, and as noted in \citet{Johannsen:2013_JP}, when expressed in this form the JP metric satisfies asymptotic flatness, recovers the correct Newtonian limit and fulfills current PPN constraints.
At the lowest order in the expansion, the JP metric depends on $\epsilon_{3}$, $\alpha_{13}$ , $\alpha_{22}$, and $\alpha_{52}$.
The Kerr metric is recovered if all of these parameters are set to zero.
As in \citet{Younsi2023}, we set $\alpha_{52}=0$, since it modifies the $g_{rr}$ component of the metric, and when fixed to zero ensures that the (outer) event horizon radius is equal to the Kerr value of $r_{\rm H} = M + \sqrt{M^{2} - a^{2}}$.
Regarding quantification of deviations from Kerr, Figs. 10 and 11 in \citet{Younsi2023} show the impact of altering JP parameters on the radii of the ISCO and the unstable photon orbit (which controls the BH shadow image size). Mathematical lower bounds on these deviation parameters are detailed in \cite{Johannsen:2013_JP}, and observational bounds on $\alpha_{13}$ at 2PN order have been found via EHT shadow size measurements \citep{Psaltis+2020}.

In practice the covariant magnetohydrodynamic (MHD) simulations of the JP metric must be performed in coordinates which remove the coordinate singularity at the (outer) event horizon \citep[e.g., see][for a discussion of horizon-penetrating coordinates]{Kocherlakota:2023}.
Following \citet{Johannsen:2013_JP}, the above line element may be expressed in horizon-penetrating form as
\begin{equation}
\begin{aligned}
{\rm d}s^{2} = &-\frac{\widetilde{\Sigma} \, \mathcal{B}}{\mathcal{F}} \, {\rm d}t^{2}
+ \frac{\widetilde{\Sigma} \, X}{\mathcal{F} \sqrt{A_{5}}} {\rm d}t \, {\rm d}r
- \frac{2 a \, \widetilde{\Sigma} \, \mathcal{C} \sin^{2} \theta}{\mathcal{F}} \,{\rm d}t \, {\rm d}\phi 
+ \frac{\widetilde{\Sigma} \, A_{1} \, Y}{\mathcal{F}\, A_{5}} \, {\rm d}r^{2} \\
&- \frac{2 a \, \widetilde{\Sigma} \, Z \sin^{2} \theta}{\mathcal{F} \sqrt{A_{5}}} \,{\rm d}r \, {\rm d}\phi 
+ \widetilde{\Sigma} \, {\rm d}\theta^{2} 
+ \frac{\widetilde{\Sigma} \, \mathcal{D} \sin^{2}\theta}{\mathcal{F}} \, {\rm d}\phi^{2} \,,
\label{JP_KS}
\end{aligned}
\end{equation}
where
\begin{subequations}
\begin{eqnarray}
X &\equiv& a^{2} A_{1} A_{2}^{2} \sin^{2}\theta + 2MrA_{1} - a^{2} A_{2} \sin^{2}\theta \,, \\
Y &\equiv& a^{2} A_{1} A_{2}^{2} \sin^{2}\theta +  2MrA_{1} - 2 a^{2} A_{2} \sin^{2}\theta + (r^{2} + a^{2})A_{1} \,, \\
Z &\equiv& 2MrA_{1} - a^{2} A_{2} \sin^{2}\theta + (r^{2} + a^{2})A_{1}^{2} \, A_{2} \,.
\end{eqnarray}
\end{subequations}
The JP metric is initialized in the horizon-penetrating (Kerr-Schild-like) form, as expressed in eqn.~\eqref{JP_KS} above, enabling the MHD evolution of the accreting plasma to flow smoothly across the event horizon and ensuring numerical stability.
In the practical numerical simulations, we also adopt a logarithmic radial coordinate in which $x^{1} \equiv \ln r$, which is standard practice in modern GRMHD simulations \citep{Porth:19} and concentrates grid resolution near the event horizon.
Radiative transfer post-processing of the MHD simulations using \texttt{BHOSS} is performed in the same coordinates as the MHD simulations.
We note that there are other metric parametrization frameworks which seek to represent BH solutions in different theories of gravity \citep[e.g.,][]{Vigeland_2011,Konoplya_2016}.

\begin{figure}
    \centering
    \includegraphics[width=\columnwidth,trim= 160pt 0pt 200pt 0pt, clip]{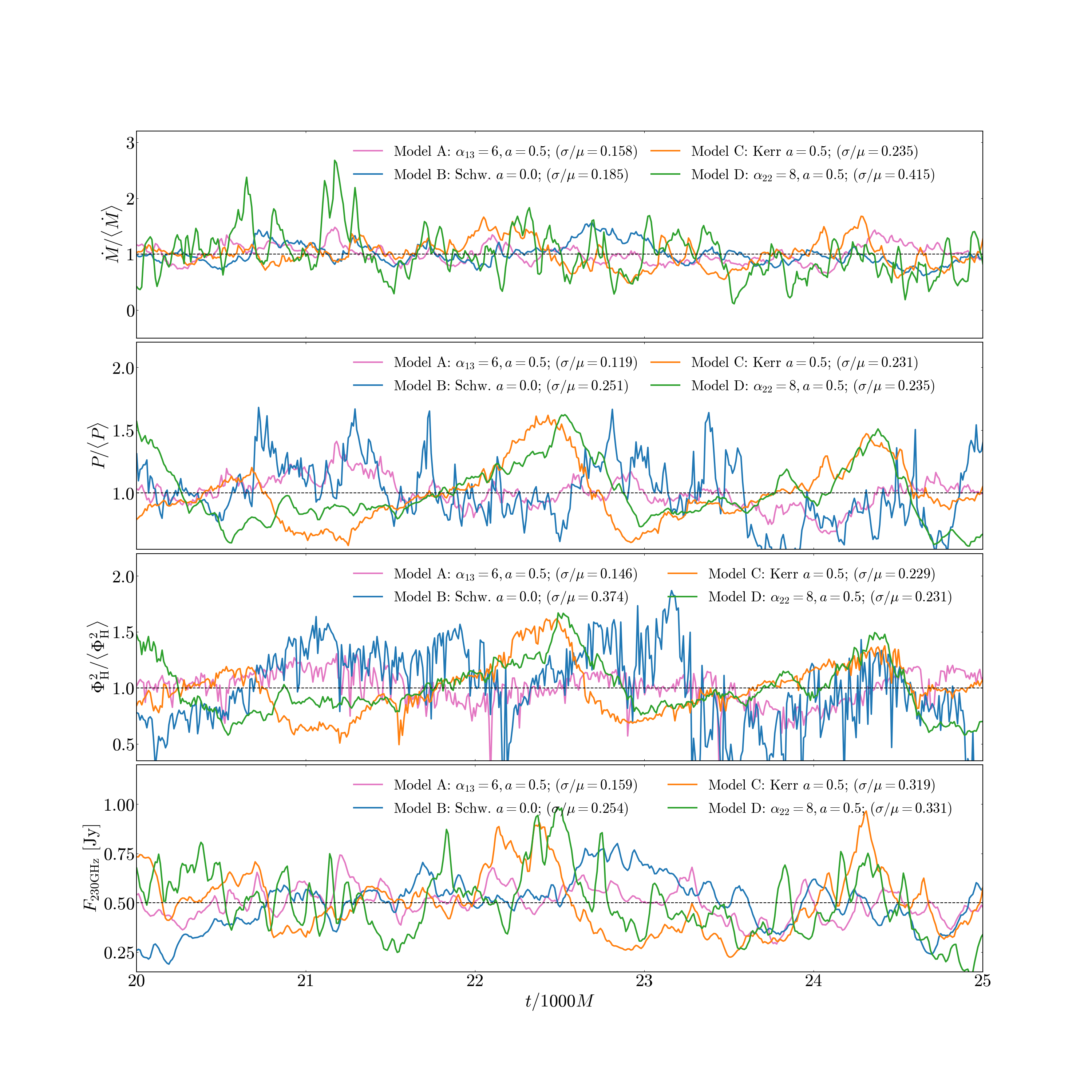}
    \caption{Time evolution of the accretion rate $\dot{M}$, the outflow power $P$, the square of the un-normalized horizon magnetic flux $\Phi_{\rm H}^2$ (which is essentially equivalent to the expected BZ power) and the 230 GHz flux for four different (non-)Kerr models.
    We also mention the ratio of the $1\sigma$ standard deviation and the mean $\mu$ for each curve as a measurement of the variability of each quantity for each model.
    }
    \label{fig:evolution}
\end{figure}

\begin{figure*}
    \centering
    \includegraphics[width=\textwidth,trim= 110pt 0pt 140pt 0pt, clip]{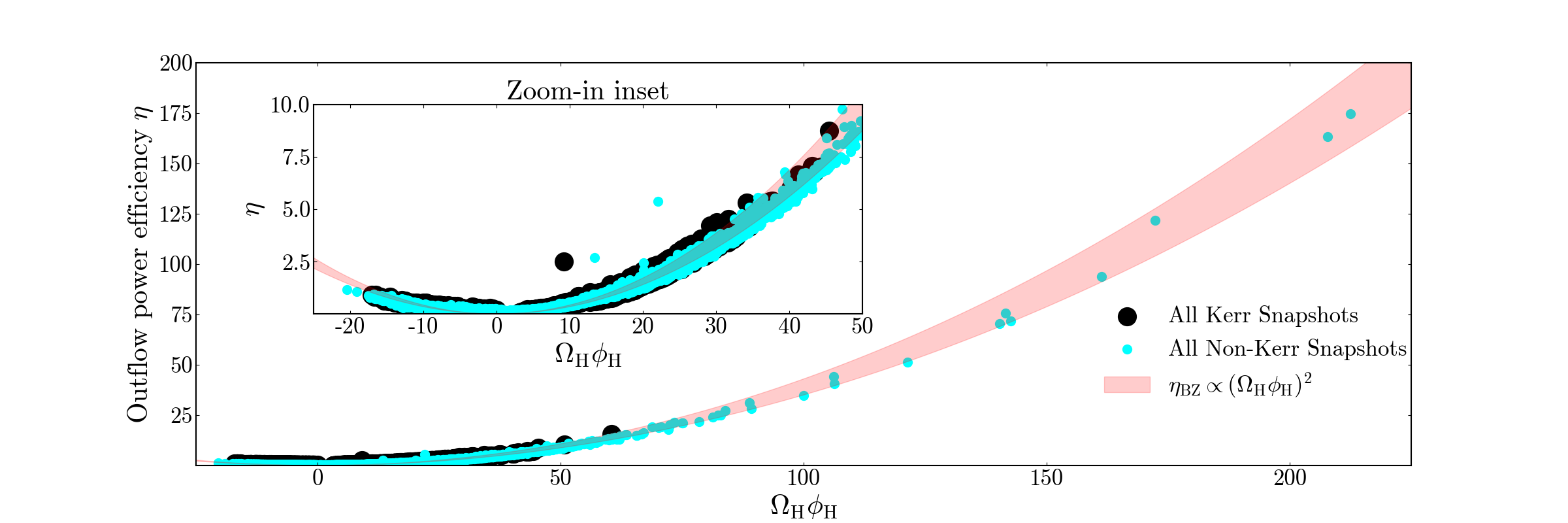}
    \caption{We find that the instantaneous outflow efficiencies from all models and their individual time snapshots fall in line with the BZ prediction. The inset shows the same figure but zoomed in to a smaller range in $\OmegaH\phibh$. Fig.~\ref{fig:BZ}(a) shows the time-averaged $\eta$ instead.
}
    \label{fig:BZ_inst}
\end{figure*}

\section{\label{sec:fluxes}Accretion flow properties and time variability}

In our main Letter, we focus on time-averaged properties of the different models, namely, their outflow efficiencies and ray-traced 230~GHz and 86~GHz images. 
However, one of the advantages of using these time-dependent models is to capture the effects of turbulence and magneto-hydrodynamic instabilities in the accretion flow and outflow evolution. 
For this, one must adequately resolve the dominant process behind disk gas turbulence, which is the magneto-rotational instability \citep[MRI;][]{bal91} in a differentially rotating magnetized flow. 
Following standard definitions in the GRMHD community \citep{Porth:19}, we measure MRI quality factors, which specify the effective number of grid cells resolving the largest MRI wavelength within the disk, weighted by gas density as a proxy for the disk bulk:
\begin{equation}
    Q_{r, \theta, \varphi} \equiv  \frac{\langle 2\pi \, v_{\rm A}^{r, \theta, \varphi}\rangle_{\rho}}{\langle\Delta^{r,\theta,\varphi} \,\omega\rangle_{\rho}}\,,
\end{equation}
where the Alfv\'en speed in the $i$-th direction is denoted $v_{\rm A}^i$, the cell size in the same direction is $\Delta^i$, and the fluid angular velocity is $\omega$.
While convergence tests for disk bulk properties such as the effective viscosity, plasma beta, and gas density distributions are still a topic of active research due to the complex feedback of the microphysics of turbulence and macrophysics of large-scale eddies, bulk inflows, and outflows, it is generally found that $Q\gtrsim 10-15$ is sufficient for broad convergence of disk parameters \citep{hgk11, Porth:19}. The time- and density-averaged $Q$ values from our models are $Q_r\approx 31$, $Q_{\theta} \approx 32$ and $Q_{\varphi} \approx 80$, which exceed the minimum requirements substantially. 

Moving on to the time-evolution of the disk, we provide the definitions of the normalized magnetic flux $\phibh \equiv \Phi_{\rm H} /\sqrt{\dot{M}}$, accretion rate $\dot{M}$, and the outflow power $P \equiv \dot{M}c^2 - \dot{E}$ in Sec.~\ref{sec:methods}. Figure~\ref{fig:evolution} shows the time dependence of $\dot{M}$, $P$ and $\Phi_{\rm H}^2$, normalized by their respective mean values, of four models over $20000$--$25000~M$. 
Of these four models, one is our Schwarzschild BH and the other three are chosen to have the same BH spin $a=0.5$, but with different non-Kerr parameters, one Kerr, one $\alpha_{13}=6$ and one $\alpha_{22}=8$ model, to show the change in accretion properties for a single BH spin value. The outflow efficiency increases from Model A to D.

The temporal profiles of each quantity for all models display considerable variability, measured by the ratio of the $1\sigma$ standard deviation and the mean $\mu$ of the corresponding time-curve. 
There are periods of increased accretion and outflow production (shown by peaks in $\dot{M}$ and $P$), which do not always correlate. 
On the other hand, the square of the horizon magnetic flux $\Phi_{\rm H}^2$ and the outflow power correlate very well since Blandford-Znajek driven jets dominate the outflow for magnetically arrested spinning BHs.
Note that $P_{\rm BZ}/\langle P_{\rm BZ} \rangle = \Phi_{\rm H}^2/\langle \Phi_{\rm H}^2 \rangle$ as all other terms are constant for a single model. 

A notable feature of magnetically arrested disks (MADs) is the production of magnetic flux eruptions via magnetic reconnection close to the BH. When the BH magnetosphere (i.e., the magnetized plasma environment around the BH) is over-saturated with magnetic flux, the system expels a large amount of this magnetic flux (seen as dips in $\Phi_{\rm H}$ in Fig.~\ref{fig:evolution}) via the ejection or ``eruption'' of bunched vertical field lines, otherwise known as a magnetic flux tube (which are also seen in other systems such as the solar corona), disrupting the incoming accretion flow and churning up further MHD instabilities such as Kelvin-Helmholtz and Rayleigh-Taylor-like instabilities \citep{Chatterjee:2022,Ripperda2022,Zhdankin2023,Chatterjee:2024}. These eruption events are responsible for most of the variability seen in MAD models.

The time-variability of the accretion rate, horizon magnetic flux and outflow power in MAD models allows us to probe accretion flows of, effectively, different magnetization parameters, including the weakly magnetized regime \citep[or ``SANE'' flows; e.g.,][]{Chatterjee:2022}. Remarkably, we find that the instantaneous outflow efficiency (Fig.~\ref{fig:BZ_inst}), calculated from $\dot{M}$ and $P$ at every time snapshot, also follows the BZ power predictions, just as shown in the main Letter for the time-averaged powers (Fig.~\ref{fig:BZ}). The instantaneous efficiency and the normalized magnetic flux can become quite large during eruption events, when $\dot{M}\rightarrow0$, enabling us to validate the BZ mechanism at much more extreme efficiencies. However, instantaneous quantities are not as useful as time-averaged ones for comparison to observations since observational data is often averaged over the viewing window and simultaneous multi-frequency data for a source are rare. Hence we choose to report time-averaged simulation results in the main Letter.

\begin{figure}
    \centering
    \includegraphics[width=\columnwidth,trim= 150pt 0pt 100pt 0pt, clip]{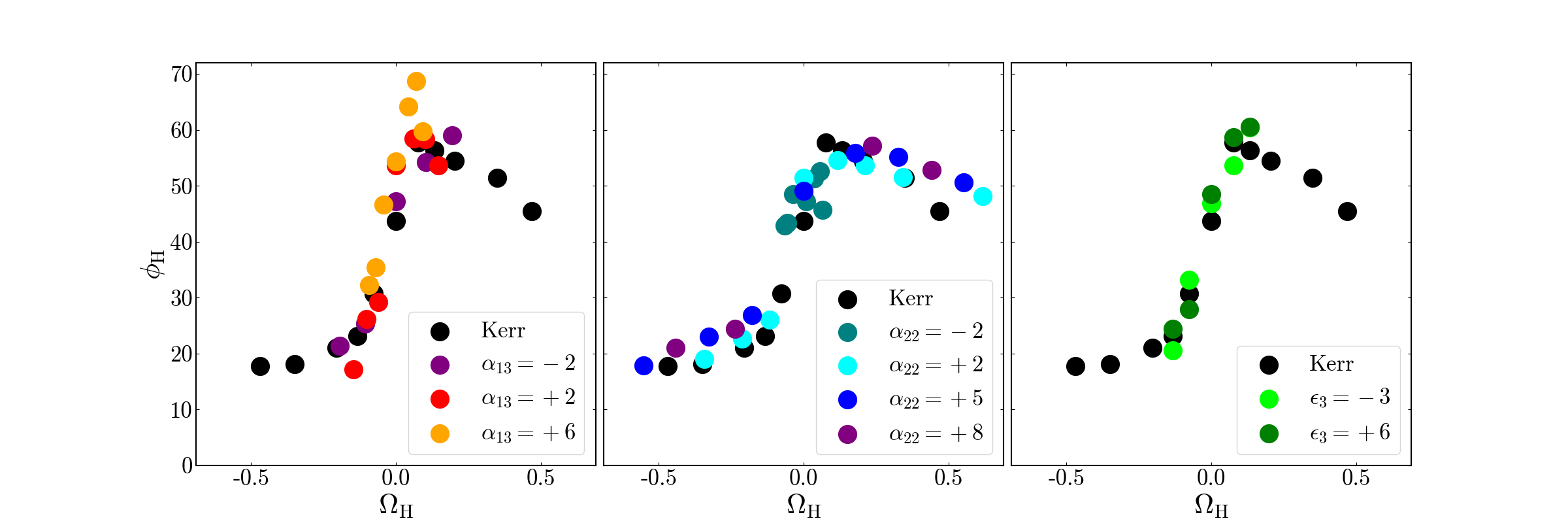}
    \caption{Accretion-rate normalized horizon magnetic flux $\phibh$ for our model set as a function of the corresponding horizon angular frequency $\OmegaH$. We find that there is an approximate cubic dependence on $\OmegaH$ with the largest deviations seen in the non-Kerr $\alpha_{13}$ models.
    }
    \label{fig:phibh}
\end{figure}

The maximum achievable $\Phi_{\rm H}$, i.e., its saturation value, is known to depend on accretion flow properties as well as the underlying spacetime, but a precise analytical description is yet to be derived.
The result should in principle be an outcome of the radial force equation, where the outward-pointing force due to magnetic fields balances the ram pressure from the gravitationally-attracted accreting gas, with complications primarily due to the non-axisymmetric structure of the inflow. 
From Fig.~\ref{fig:evolution} there is a clear indication that models with larger outflow powers exhibit larger variability in flow properties, suggesting that the energy outflow from the BH magnetosphere could dictate, or at least indirectly affect, the frequency and morphology of magnetic eruptions.
Further investigation is necessary to reveal causal factors behind determining the saturated $\Phi_{\rm H}$, which has been shown to follow a cubic dependence on BH spin for Kerr BHs \citep{tch12proc, Narayan:2022} and an approximate cubic dependence on the horizon angular frequency $\OmegaH$ for generalized BH models in this work (Fig.~\ref{fig:phibh}). 

We also show the 230 GHz lightcurves for each model and their variability. Generally the lightcurves follow variations in the accretion rate, with some contribution from the corresponding jets. In future work, we will tackle lightcurves and their properties in more detail, with greater exploration of the radiative properties of the plasma.

\begin{figure}
    \centering
    \includegraphics[width=\columnwidth,trim= 0pt 0pt 0pt 0pt, clip]{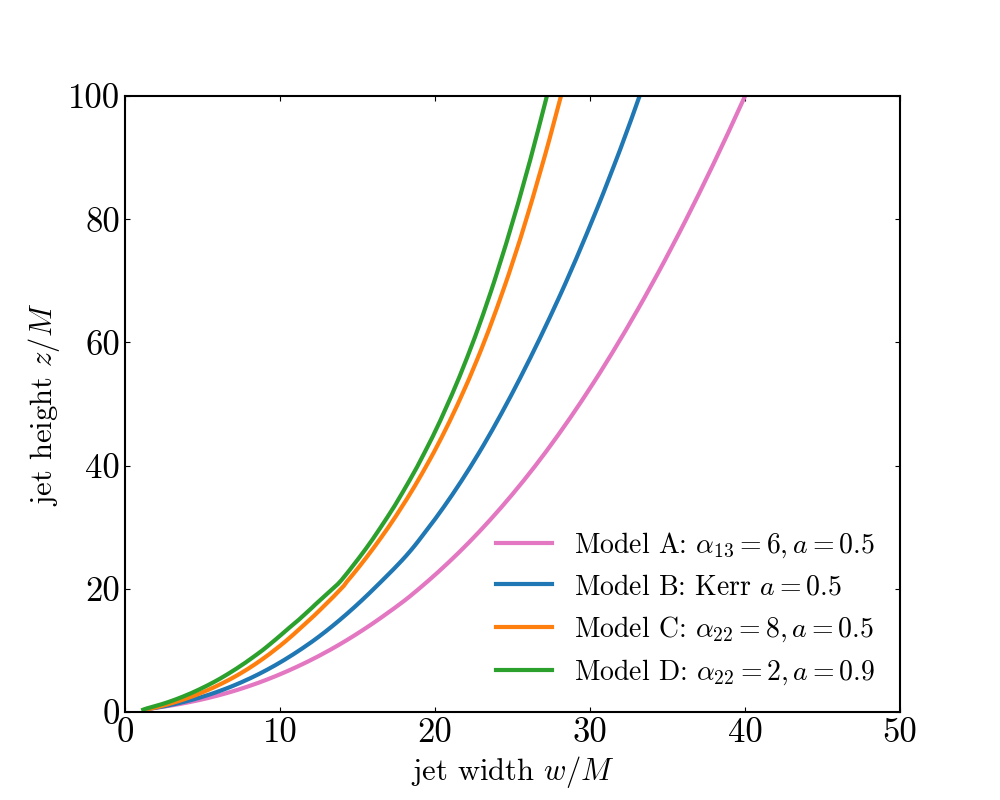}
    \caption{Jet shape profiles for four models with jets. The horizon angular frequency $\OmegaH$ and the jet efficiency $\eta$ both increase as one moves from Model A through to D.
    It is found that the jet becomes increasingly more collimated as its power increases. 
    }
    \label{fig:jetopening}
\end{figure}

\section{\label{sec:collimation}Jet properties: collimation}

Our main Letter deals with energy extraction from BHs described by spacetime metrics which parametrically deviate from the standard Kerr BH. 
We find that the jet power predicted from the Blandford-Znajek mechanism \citep[BZ;][]{bz77} accurately describes the energy outflow seen in our models, emphasizing the generality of frame-dragging and magnetic field coupling around BHs. 
As remarked upon in the Letter, increasing the horizon angular frequency, either by increasing the BH spin or the BH $\alpha_{22}$ parameter, dramatically changes the jet power as well as the horizon-scale BH image properties, such as the photon ring shape and the shadow size.
At larger distances from the BH, the collimation profile of the radio jet, such as the 86~GHz images shown for M87$^*$, is a direct consequence of the collimation of the underlying physical jet. 
As $\OmegaH$ increases, the magnetic field lines in the BH ergosphere become more strongly twisted, causing the jet to collimate into a narrower structure. Figure~\ref{fig:jetopening} shows the jet collimation profiles from four GRMHD simulations where $\OmegaH$ increases from model A to model D. 

Jets have high magnetic energies compared to the rest-mass energy of the gas and so the jet boundary is defined as the surface where the magnetic and rest-mass energies are equal (i.e., the magnetization is unity). 
Parsec/kiloparsec-scale jets from active galactic nuclei (AGN) exhibit a wide range of jet shapes, ranging from parabolic to conical profiles: $\lambda\approx 0.39-0.56$ \citep{Kovalev2020} assuming that the jet length $z$ and its width $w$ are related as $w\propto z^{\lambda}$. 
The power-law index $\lambda$ for our models falls within the range shown by AGN jets and agrees with estimates from other GRMHD codes \citep{Narayan:2022}.
We also note that weakly-powered jets ($\eta\lesssim 10\%$) also appear considerably collimated. 
The primary difference is that weakly-powered jets are collimated by their environment (comprising disk winds and interstellar medium gas) whereas strongly-powered jets generally overcome the ram pressure from their surrounding gas, with their collimation profile dependent on how much the horizon-scale magnetic field is twisted by frame-dragging. 

\bibliography{Refs-JP-GRMHD}{}
\bibliographystyle{aasjournalv7}



\end{document}